\documentclass[draft,grl]{AGUTeX}

\usepackage{graphicx}

\setkeys{Gin}{draft=false}

\authorrunninghead{ORUBA AND DORMY}

\titlerunninghead{DIPOLAR - MULTIPOLAR DYNAMOS TRANSITION}

\authoraddr{Ludivine Oruba, Emmanuel Dormy, 
MAG (ENS/IPGP), 
LRA, D\'epartement de Physique, 
Ecole Normale Sup\'erieure,
24, rue Lhomond,
75231 Paris Cedex 05, France
(loruba@lra.ens.fr, dormy@phys.ens.fr)}

\begin{document}

%
%
%
%
%

%
%

\title{Transition between viscous dipolar and inertial multipolar dynamos}

%
%

\authors{Ludivine Oruba, \altaffilmark{1}
Emmanuel Dormy, \altaffilmark{1,2}}

\altaffiltext{1}
{MAG (ENS/IPGP)\break
LRA, D\'epartement de Physique\break
Ecole Normale Sup\'erieure\break
24, rue Lhomond\break
75231 Paris Cedex 05, France.}

\altaffiltext{2}{Institut de Physique du Globe de Paris, CNRS/UMR7154.}

%
%


\begin{abstract}









We investigate the transition from steady dipolar to reversing multipolar
dynamos. The Earth has been argued to lie close to this transition, which
could offer a scenario for geomagnetic reversals.
We show that the transition between dipolar and 
multipolar dynamos
is characterized by a three terms balance (as opposed to the usually
assumed two terms balance), which involves the non-gradient parts of inertial, 
viscous and Coriolis forces.
We introduce from this equilibrium the sole parameter 
${{\rm Ro}}\,{{\rm E}}^{-1/3} \equiv {{\rm Re}}\,{{\rm E}}^{2/3}$, which accurately describes the
transition for a wide database of 132 fully three dimensional direct
numerical simulations of spherical rotating dynamos (courtesy of U. Christensen). 
This resolves earlier contradictions in the literature on the relevant two
terms balance at the transition.
Considering only a two terms balance between the non-gradient part of the
Coriolis force and of inertial forces, provides the classical
${{\rm Ro}}/{\ell_u}$ (Christensen and Aubert, 2006).
This transition can be equivalently described by ${{\rm Re}} \, {\ell^{2}_u}$,
which corresponds to the two terms balance between the non-gradient part 
of inertial forces and viscous forces (Soderlund {\it et al.}, 2012).

\end{abstract}

%
%

%

\begin{article}

\section{Introduction}

The transition between the dipolar dynamo regime and a non-dipolar regime
was first pointed out by Kutzner and Christensen in 2002. They observed through direct
numerical experiments two different regimes of dynamo action. One, at
moderate forcing, characterized by a steady large scale dipole and the
other, with a more vigorous forcing, revealing multipolar solutions and
chaotic reversals of the dipolar component.
The existence of these two regimes and their possible relevance to the
geodynamo is a central problem in geomagnetism.

Christensen and Aubert (2006) found through a wide parameter space survey
that this transition is controlled by the relative strengths of inertial  
and Coriolis forces, as measured by the so called ``local Rossby number''. 
More recently, Soderlund {\it et al.} (2012) argued that the Coriolis force was
dominant in both regimes, and
suggested, instead, that the transition was controlled by the relative strengths of 
inertial to viscous forces, both of lower amplitude.

In this letter, we use 
a wide database
of 132 fully three dimensional direct numerical simulations 
(kindly provided by U. Christensen) to
address this apparent contradiction. We argue that both interpretations are in fact equivalent if
one considers the non-gradient part of the forces balance only. This is
easily achieved by considering the curl of the relevant forces. The fluid
being incompressible, the gradient part of any force will obviously be balanced by
pressure forces. 

\section{Modeling}

Our study relies on a database of numerical simulations
performed in a spherical shell of typical width $L=r_o-r_i$ and aspect ratio $\xi\equiv r_i/r_o=0.35 \,$. The 
flow is thermally driven by an imposed difference of temperature between the inner and outer spheres. These 
simulations rely on no-slip mechanical boundary
conditions, and an insulating outer domain. Most of simulations involve an insulating inner core, a few of them 
involve a conducting inner core with the same conductivity as the fluid.

The governing equations in the rotating reference frame can then be
written in their non-dimensional form -- using $L$ as unit of length, $(2\Omega)^{-1}$ as unit of
time, $\Delta T$ as unit of temperature, and $ 2\,
\Omega \, \sqrt{\rho \mu} \, L$ as
unit for the magnetic field -- as
\begin{equation}
\partial _t {\mbox{\bf u}} +  ({\mbox{\bf u}} \cdot {\mbox{\boldmath $\nabla$}}) {\mbox{\bf u}}
=
- {\mbox{\boldmath $\nabla$}} \pi
+ {\rm E} \, \Delta {\mbox{\bf u}}
- {\mbox{\bf e}}_z \times {\mbox{\bf u}} 
+ \frac{{\rm Ra}\, {\rm E}^2}{\rm Pr} \, T \, \frac{{\mbox{\bf r}}}{r_o}
+ \left({\mbox{\boldmath $\nabla$}} \times {\mbox{\bf B}} \right) \times {\mbox{\bf B}}\, , \,\,\,\,\,\,
\label{eq_NS}
\end{equation} 
\begin{equation}
\partial _t {\mbox{\bf B}} = {\mbox{\boldmath $\nabla$}} \times ({\mbox{\bf u}} \times {\mbox{\bf B}}) 
+ \frac{\rm E}{\rm Pm} \, \Delta {\mbox{\bf B}} \, ,
\qquad
\partial_t T + ({\mbox{\bf u}} \cdot {\mbox{\boldmath $\nabla$}}) T
= \frac{\rm E}{\rm Pr} \Delta T\, ,
\label{eq_ind}
\end{equation} 
\begin{equation}
{\mbox{\boldmath $\nabla$}} \cdot {\mbox{\bf u}} = 
{\mbox{\boldmath $\nabla$}} \cdot {\mbox{\bf B}} = 
0\, .
\label{eq_div}
\end{equation} 

These equations involve four non-dimensional numbers:
the Ekman number ${{\rm E}} = \nu / (2\Omega L ^2) \, ,$
the Prandtl number ${\rm Pr}={\nu}/{\kappa}\, , $ 
the magnetic Prandtl number ${\rm Pm}={\nu}/{\eta}\, ,$
and the Rayleigh number ${{\rm Ra}} = \alpha g_0 \Delta T L ^3 / (\nu \kappa)\, ,$
in which $\nu$ is the kinematic viscosity of the fluid, 
$\alpha$ the coefficient of thermal expansion, $g_0$ the gravity at the outer bounding sphere, 
$\kappa$ its thermal diffusivity, 
and $\eta$ its magnetic diffusivity. 

The database used for this study covers the parameter range 
\({{\rm E}} \in [5\times10^{-7},\,5 \times 10^{-4}]\)
and \(\, {{\rm Pm}} \in [0.04,\, 33.3] \), and is restricted to \({{\rm Pr}} =1 \, .\)
We therefore cannot distinguish any possible dependence on the Prandtl number 
${{\rm Pr}}$ from our analysis.

\section{Coriolis versus inertial forces}
\label{physic1}
\begin{figure}
\centerline{\includegraphics[width=1\textwidth]{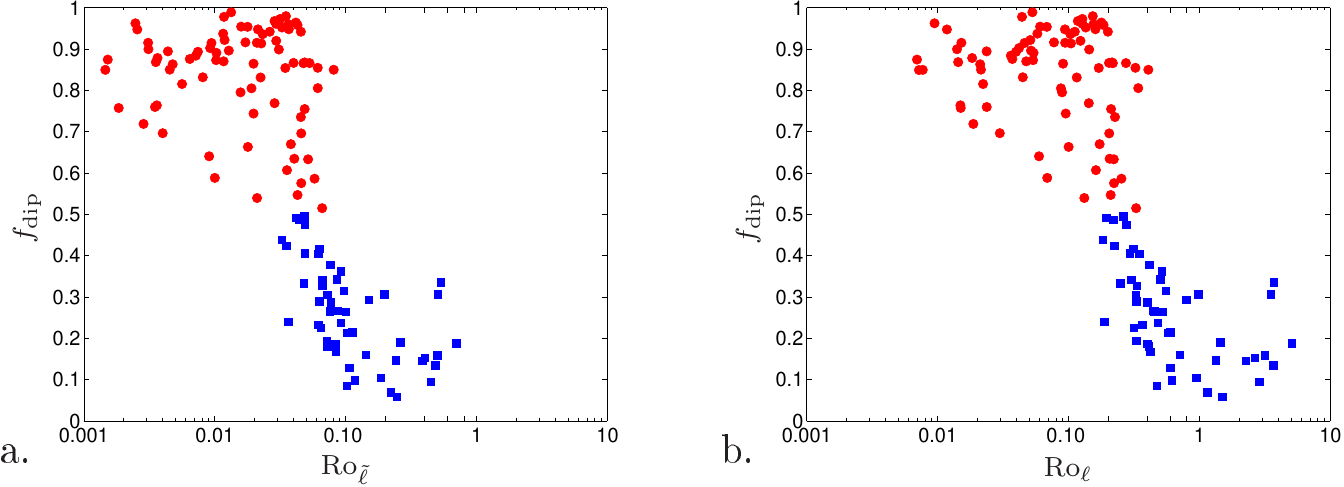}}
\caption{ \normalsize Dipolarity as measured by ${{f_{\rm dip}}}$ versus (a) the local Rossby number ${{\rm Ro}}_{\tilde{\ell}}$ as defined by 
Christensen and Aubert (2006), (b) the local Rossby number ${{\rm Ro}}_{\ell}$ as defined by Oruba and Dormy (2014). 
Red circles (resp. blue squares)
 correspond to ${{f_{\rm dip}}}>0.5$ (resp. ${{f_{\rm dip}}}<0.5$). These graphs rely on 
the $132$ dynamos database.}
\label{fdip_Ro}
\end{figure}

The dipolarity of the magnetic field is well quantified by ${{f_{\rm dip}}}$, which corresponds to the time-averaged ratio 
of the mean dipole field strength to the field strength in harmonic 
degrees $n=1-12$ at the outer bounding sphere (see Christensen and Aubert, 2006). We
also introduce the Rossby number ${{\rm Ro}}$, defined using time-averaged quantities as 
\begin{equation}
{{\rm Ro}} \equiv \langle {{\mbox{\bf u}}}^2\rangle^{1/2} \equiv {{\rm u}} \, ,
\end{equation} 
where $\langle . \rangle$ denotes the volume average over the shell.
Christensen and Aubert (2006) empirically show that the transition 
from the dipolar regime to the non-dipolar regime occurs at ${{\rm Ro}}_{\tilde{\ell}}\simeq 0.1$ (which becomes 
${{\rm Ro}}_{\tilde{\ell}}\simeq 0.05$ with our definition of the time scale), where ${{\rm Ro}}_{\tilde{\ell}}$ denotes 
the local Rossby number, defined as ${{\rm Ro}}_{\tilde{\ell}}={{\rm Ro}}/\tilde{\ell}_u$. 
This transition was also shown to work with stress free boundary conditions (Schrinner {\it et al.}, 2012). 
The typical length scale of the flow $\tilde{\ell}_u$ 
is here defined as $\pi/\overline{n}$, where $\overline{n}$ corresponds to the mean value 
of the spherical harmonics degree $n$ in the time-averaged kinetic
energy spectrum (see equation~(27) in Christensen and Aubert, 2006).  
Figure~\ref{fdip_Ro}.a represents the dipolarity ${{f_{\rm dip}}}$ as a function of ${{\rm Ro}}_{\tilde{\ell}}$, when applied 
to the $132$ dynamos database. 

The above definition of $\tilde{\ell}_u$ is anisotropic as it does not take the 
radial variations into account. Indeed it only involves the spherical 
harmonics degree $n$ of the spectral decomposition. Moreover, this
calculus relies on a radial averaging: the role of the radius $r$ in the
length scale associated to a given angular  
scale is thus not taken into account. 
Oruba and Dormy (2014) introduced the kinematic dissipation length scale, denoted 
here as $\ell_u$, defined using time-averaged quantities 
as
\begin{equation} 
\ell_u ^2\equiv \frac{\langle \mathbf{u}^2 \rangle}{\langle ({\mbox{\boldmath $\nabla$}} \times \mathbf{u})^2 \rangle} \, .
\label{ldissu}
\end{equation} 
Contrary to $\tilde{\ell}_u$, this length scale ${\ell}_u$ is isotropic.
Using the $132$ dynamos database, the two length scales are found roughly proportional, with 
$\tilde{\ell}_u \simeq 5 {\ell}_u$~: as expected $\tilde{\ell}_u > {\ell}_u$. 
We can then introduce the local Rossby number ${{\rm Ro}}_{\ell}$, defined as 
${{\rm Ro}}_{\ell}={{\rm Ro}}/\ell_u$. The transition between the dipolar regime and the multipolar regime 
occurs for ${{\rm Ro}}_{\ell}$ closer to unity (see figure~\ref{fdip_Ro}).

The role of the local Rossby number in the dipolar-multipolar
transition obviously relates to a dominant forces balance between inertial 
and Coriolis
forces in the Navier-Stokes equation~(\ref{eq_NS}).
This transition is associated with 
a breakdown of the dominant dipolarity of the dynamo as inertia increases, 
but it happens when inertial forces are still too small to suppress the 
columnar 
structure of convection. This subsequent hydrodynamic breakdown occurs 
at larger forcing and is discussed in Soderlund {\it et al.} (2013).

The transition between the dipolar regime and the multipolar regime is then
expected to occur when the non-gradient part of inertial forces is 
of the same order of magnitude as that of the Coriolis force. 
This statistical equilibrium can be expressed by considering the curl of this
forces balance, it provides
\begin{equation} 
\left\{{\mbox{\boldmath $\nabla$}} \times \left({\mbox{\bf u}} \times {\mbox{\boldmath $\nabla$}} \times {\mbox{\bf u}} \right) \right\} \sim 
\left\{{\mbox{\boldmath $\nabla$}} \times \left({\mbox{\bf e}}_z \times {\mbox{\bf u}} \right) \right\}     \,, 
\end{equation} 
which can be rewritten as 
\begin{equation}
\frac{{{{\rm u}}}^2}{\ell \, \ell_u} \sim
\frac{{{\rm u}}}{\ell_{/\!/}}\, .
\end{equation} 
Clearly $\ell_u$, as occuring in this relation, corresponds to the dissipation length scale 
defined in (\ref{ldissu}) and not to the often used $\tilde{\ell}_u$. 
In the following, we will thus only consider the length scale ${\ell}_u$. 
Below and at the transition, it is natural to assume
\(\ell_{/\!/} \sim 1\).
This implies
\begin{equation} 
{{\rm Ro}}_{\ell} \sim \ell\,,
\label{equil_inertieCor}
\end{equation} 
where $\ell$ is a non-dimensional length scale which depends on correlations 
between the velocity and the vorticity. This length scale is thus an intricate
quantity, which cannot be easily estimated a priori. 

Note that assuming that $\ell=1$ is tantamount to writing the balance
between inertial forces and the Coriolis force  
without taking the curl of each term. 

\section{Coriolis versus viscous forces}

King and Buffett (2013) have shown that in numerical simulations, 
the typical length scale of the flow ${\tilde{\ell}_u}$ verifies ${\tilde{\ell}_u} \sim {{\rm E}}^{1/3}$. The scaling is even more naturally satisfied by the  dissipation length scale $\ell_u \, .$

\begin{figure}
\centerline{\includegraphics[width=0.5\textwidth]{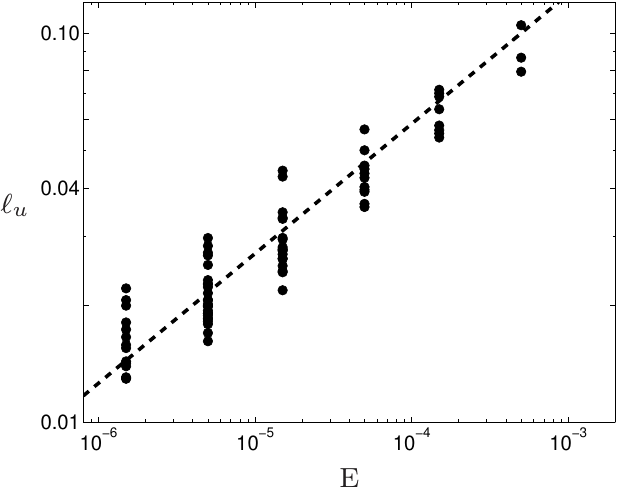}}
\caption{The non-dimensional characteristic length scale $\ell_u$ as a function of the Ekman number.
The dashed line corresponds to $\ell_u \sim {{\rm E}}^{1/3}$. 
This plot relies on a subset of $81$ dynamos, corresponding to ${{f_{\rm dip}}}>0.5$.}
\label{luE13}
\end{figure}

The $E^{1/3}$--scaling stems from the equilibrium between the non-gradient part of the viscous force and the Coriolis force
\begin{equation} 
\left\{{{\rm E}} \, {\mbox{\boldmath $\nabla$}} \times {\mbox{\boldmath $\nabla$}} \times {\mbox{\boldmath $\nabla$}} \times {\mbox{\bf u}} \right\}  
\sim
\left\{ {\mbox{\boldmath $\nabla$}} \times \left({\mbox{\bf e}}_z \times {\mbox{\bf u}} \right) \right\} 
\,, 
\end{equation}
which provides
\begin{equation}
\frac{{{\rm E}} \, {{{\rm u}}}}{ {\ell_u}^3} 
\sim
\frac{{{\rm u}}}{\ell_{/\!/}}
\, .
\end{equation} 
This implies ${\ell}_u \sim {{\rm E}}^{1/3}$ which is well verified 
by dipolar dynamos (${{f_{\rm dip}}}>0.5$) in the database (see figure~\ref{luE13}). 

\section{Inertial versus viscous forces}

Because of this dominant balance between viscous forces and the Coriolis term, valid in
the dipolar regime, the transition to multipolar dynamos necessarily also
corresponds to a balance between inertial and viscous forces.

\begin{figure}
  \centerline{\includegraphics[width=1\textwidth]{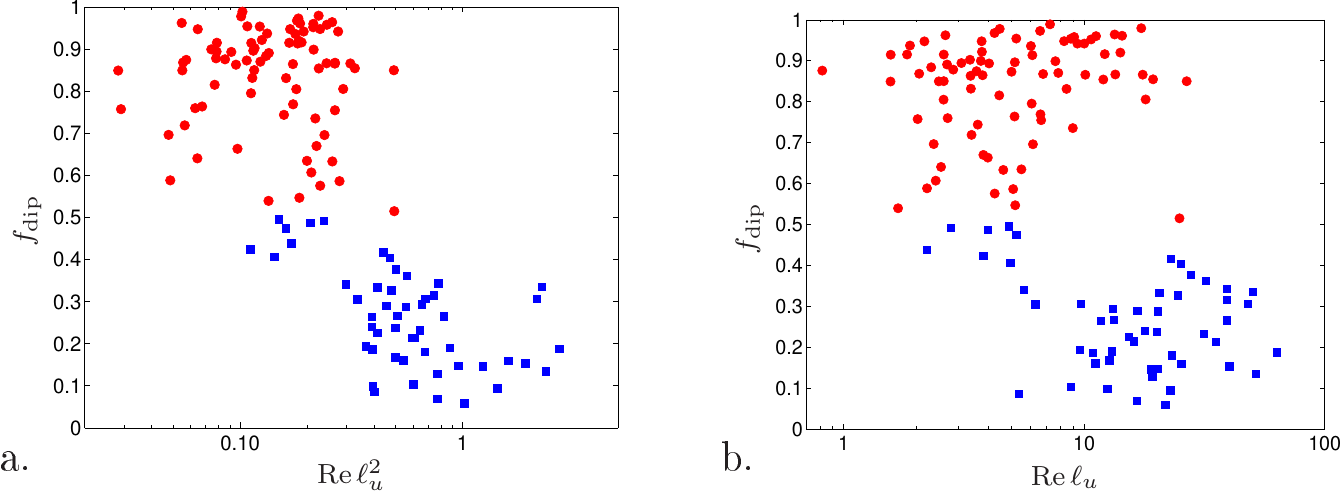}}
\caption{Dipolarity as measured by ${{f_{\rm dip}}}$ versus (a) ${{\rm Re}} \, {\ell^{2}_u}$ and (b)${{\rm Re}} \, {\ell_u}$. Red circles (resp. blue squares)
 correspond to ${{f_{\rm dip}}}>0.5$ (resp. ${{f_{\rm dip}}}<0.5$). These 
graphs rely on the $132$ dynamos database. This figure stresses the importance of 
considering the curl of the forces balance in order to filter out the gradient parts in 
equilibrium with pressure forces.}
\label{fdip_Rel2}
\end{figure}

The statistical equilibrium between the non-gradient part of inertial and
viscous forces provides  
\begin{equation} 
\left\{{\mbox{\boldmath $\nabla$}} \times \left({\mbox{\bf u}} \times {\mbox{\boldmath $\nabla$}} \times {\mbox{\bf u}} \right) \right\} \sim 
 \left\{{{\rm E}} \, {\mbox{\boldmath $\nabla$}} \times {\mbox{\boldmath $\nabla$}} \times {\mbox{\boldmath $\nabla$}} \times {\mbox{\bf u}} \right\}     \,, 
\end{equation} 
which can be rewritten as 
\begin{equation} 
\frac{{{{\rm u}}}^2}{\ell \, {\ell_u}} \sim \frac{{{\rm E}} \, {{\rm u}}}{{\ell_u}^3}\,
\end{equation} 
and corresponds to
\begin{equation} 
{{\rm Re}} \, {\ell^{2}_u} \sim \ell\,, 
\label{equil_inertievisc}
\end{equation} 
where ${{\rm Re}}$ is the Reynolds number (${{\rm Re}} \equiv {{\rm Ro}} {{\rm E}}^{-1}$).
Figure~\ref{fdip_Rel2}.a supports the above relation.
It is interesting to note that the parameter on the left of (\ref{equil_inertievisc}) 
is not the local Reynolds number (${{\rm Re}} {\ell_u}$). Relation (\ref{equil_inertievisc}) relies on the curl of the forces balance,
and contrasts with a transition corresponding to 
a simple balance between inertial and viscous forces.
Such a naive balance would not filter the gradient parts and would yield
the local Reynolds number \({{\rm Re}} \, {\ell_u}\) as relevant parameter.
Such a scenario is obviously not supported by the numerical database (see
figure~\ref{fdip_Rel2}.b).
Taking the curl of the Navier-Stokes equation~(\ref{eq_NS}) 
is therefore necessary before writing balances between these forces, in
order to filter out the gradient part of the dominant forces.

\section{A three forces balance}

It is quite clear from the above discussion that the transition between
dipolar and multipolar dynamos occurs when the curl of inertial forces becomes
comparable to both the curl of the Coriolis and of the viscous term, which 
were both in balance in the dipolar regime. As the Rayleigh number is 
increased, the role of inertia becomes more and more important, until the three forces balance is reached.

\begin{figure}
\centerline{\includegraphics[width=0.5\textwidth]{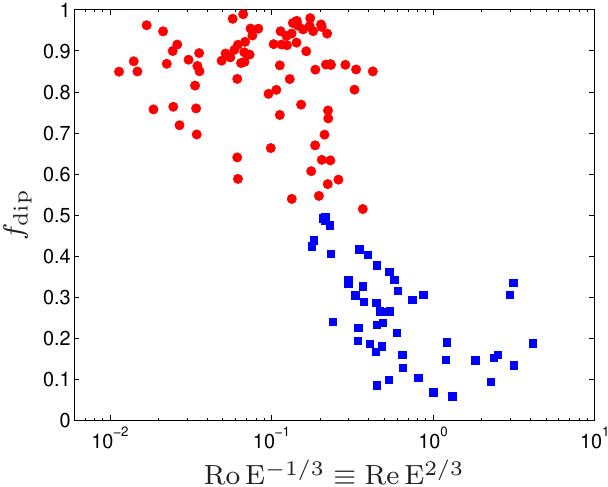}}
\caption{Dipolarity as measured by ${{f_{\rm dip}}}$ versus the sole parameter ${{\rm Ro}}\,
  {{\rm E}}^{-1/3} \equiv {{\rm Re}} \, {{\rm E}}^{2/3}$, revealing the three forces balance
  at the transition. Red circles (resp. blue squares)
  correspond to ${{f_{\rm dip}}}>0.5$ (resp. ${{f_{\rm dip}}}<0.5$). 
This figure relies on the $132$ dynamos database.}
\label{fig4}
\end{figure}

Indeed, replacing ${\ell_u}$ by ${{\rm E}}^{1/3}$ in (\ref{equil_inertieCor}) and 
(\ref{equil_inertievisc}) yields the sole relation
\begin{equation} 
{{\rm Ro}}\,{{\rm E}}^{-1/3} \equiv {{\rm Re}}\,{{\rm E}}^{2/3} \sim {\ell} \,. 
\label{paraunique}
\end{equation} 
This expression reveals the existence of a single parameter to describe
the transition. We represent the dipolarity as a function of this parameter in figure~\ref{fig4}. This parameter provides a remarkable description 
of the transition. This vindicates the above scenario with a dominant balance between three terms at the transition, that is 
the non-gradient part of inertial, viscous and Coriolis forces. 

\section{The geodynamo}
It is important to ponder on the applicability of the above transition to the
Earth's core, as this can shed light on the relevance of available
numerical models to geophysical observations. 
We use the typical estimates for the Earth's core ${{\rm E}}=10^{-14}$ 
(Olson, 2007) 
and ${{\rm Ro}}=3 \times 10^{-6}$, which corresponds to a 
typical rms velocity in the core of $1 {\rm mm}\cdot {\rm s}^{-1}$ (order
of magnitude obtained  by inverting secular variation data, see Bloxham 
{\it et al.}, 1989, and Christensen and Aubert, 2006).  
A direct estimate of the parameter 
${{\rm Ro}}\,{{\rm E}}^{-1/3} \equiv {{\rm Re}} \, {{\rm E}}^{2/3}$ then provides a typical value of $10^{-1}$. 
This is in agreement with the previous work of Christensen (2010), which
argued that the Earth's core would lie below, but close to the
dipolar-multipolar transition. The vicinity to the critical value for
transition to the multipolar state has been argued to be a possible reason for
reversals of the Earth's magnetic field (Christensen, 2010; Wicht and Tilgner, 2010).

It is important to note however that the resulting viscous length scale
would be extremely small, less than \(100\) m. It is difficult to imagine
that quasi-geostrophic structures would be stable in the Earth's core on such small
length scales. The viscous balance advocated here is usually ruled out for
the Earth's core, for which a magnetostrophic balance (between the
non-gradient part of the Coriolis and the Lorentz terms) is sought
(see for a detailed discussion Oruba and Dormy, 2014).
According to figure~\ref{fig4}, such models extended to the Earth's core
would produce a dynamo dominated by a strong dipole, yet close enough to
the multipolar region to exhibit reversals (see also Christensen, 2010).
Presently available numerical models however appear to rely on a dominant
forces balance, involving viscous forces, which is not relevant to the Earth's
core. Magnetostrophic numerical models still need to be produced.

\section{Conclusions}

We offer a unified description of the dominant forces balances at work in
numerical dynamos and of the transition between the regime of steady dipolar
dynamos and that of multipolar fluctuating dynamos. We show that it corresponds 
to a balance between three forces, i.e. 
the non-gradient part of inertial, viscous and Coriolis forces. 
This balance can be estimated by taking the curl of the dominant forces.
We derive from this three terms equilibrium the parameter ${{\rm Ro}}\,{{\rm E}}^{-1/3} \equiv {{\rm Re}}\,{{\rm E}}^{2/3}$ 
which provides an accurate description of the transition. 

Using a measured length scale in the numerical models, the transition can
be equivalently described by  ${{\rm Ro}}/{\ell_u}$ (resp. ${{\rm Re}} \,
{\ell^{2}_u}$),  
which correspond to the two forces balance between the non-gradient part of inertial (resp. viscous) and Coriolis forces. 
The transition occurs when these parameters are order one, i.e. the forces
are of comparable amplitude. 

A more accurate description of the transition would require the knowledge
of the correlation length 
scale $\ell$ which is complex to determine. Figure~\ref{fig4} indicates
that this non-dimensional length scale is of order unity in the numerical database 
used in our study. It may
be a function of other parameters (e.g. the Prandtl number {{\rm Pr}} \  which is
kept constant here). Further analysis of this correlation length scale in
direct numerical simulations could provide a finer description for the
transition.

%
%

\section*{Acknowledgments} 
The authors wish to thank Ludovic Petitdemange,
Jonathan Aurnou and Krista Soderlund for
stimulating discussions. 
The database used in this work was kindly provided by Pr.~U.~Christensen
(christensen@mps.mpg.de). 

%
%

\end{article}

\end{document}